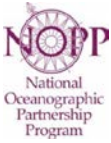 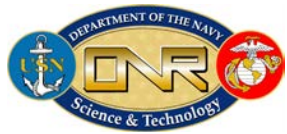 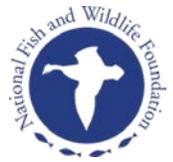

# DCL System Research Using Advanced Approaches for Land-based or Ship-based Real-Time Recognition and Localization of Marine Mammals

**Peter J. Dugan**

Bioacoustics Research Program,
Cornell Laboratory of Ornithology
Cornell University
159 Sapsucker Woods Road, Ithaca, NY 14850

phone: 607.254.1149     fax: 607.254.2460     email: pjd78@cornell.edu

**Christopher W. Clark**

Bioacoustics Research Program,
Cornell Laboratory of Ornithology
Cornell University
159 Sapsucker Woods Road, Ithaca, NY 14850

phone: 607.254.2408     fax: 607.254.2460     email: cwc2@cornell.edu

**Yann André LeCun**

Computer Science and Neural Science
The Courant Institute of Mathematical Sciences
New York University
715 Broadway, New York, NY 10003, USA

phone: 212.998.3283     mobile Phone: 732.503.9266     email: yann@cs.nyu.edu

**Sofie M. Van Parijs**

Northeast Fisheries Science Center, NOAA Fisheries
166 Water Street, Woods Hole, MA 02543
Woods Hole Oceanographic Institute, NOAA

phone: 508.495.2119     fax: 508.495.2258     email: sofie.vanparijs@noaa.gov

## LONG-TERM GOALS

We aim to investigate advancing the state of the art of detection, classification and localization (DCL) in the field of bioacoustics. The two primary goals are to develop transferable technologies for detection and classification in: (1) the area of advanced algorithms, such as deep learning and other methods; and (2) advanced systems, capable of real-time and archival and processing. This project will focus on long-term, continuous datasets to provide automatic recognition, minimizing human time to annotate the signals. Effort will begin by focusing on several years of multi-channel acoustic data collected in the Stellwagen Bank National Marine Sanctuary (SBNMS) between 2006 and 2010. Our efforts will incorporate existing technologies in the bioacoustics signal processing community, advanced high performance computing (HPC) systems, and new approaches aimed at automatically detecting-classifying and measuring features for species-specific marine mammal sounds within passive acoustic data.

## OBJECTIVES

This project represents a high-level, integrative 'bench mark' study aimed at automated detection-classification of acoustic objects for various marine mammals in a variety of ocean areas. This effort will focus on an existing acoustic dataset located in the SBNMS area, collected during an earlier National Oceanic Partnership Program (NOPP) grant. This current work will focus on investigating basic and applied acoustic detection-classification research, converting data products into white papers, reports and professional publications. We will also identify key technology areas, developing concepts for real time detection, classification and localization. In addition to publications, outputs from this work will include working prototypes and tools; developed using a maturity model based on DARPA 6.1 and 6.2 processes, for basic and applied work respectively.

## APPROACH

This effort will develop advanced methods for exploring passive acoustic data, specifically new approaches for detection-classification (deep learning) and advanced technology (high performance computing). To facilitate this work, Cornell has assembled an Integrated Research Team (IRT) of scientists, biologists and engineers, Table 1. Members of this work are highly qualified research professionals, with experience ranging from acoustical engineering, signal processing and machine learning to biology. Coordinating the efforts for this work is, (PI) Dr. Peter Dugan and, (co-PI), Dr. Christopher Clark. Joined on the team, is (co-PI) Dr. Yann LeCun, New York University (NYU) and (co-PI) Dr. Sofie Van Parijs, Northeast Fisheries Science Center (NEFSC), Woods Hole. Specialized talents for this research are broken down into three main groups. Dr. Yann LeCun and various academics at New York University, with a focus on basic research and development for applying deep learning technologies to detect and classify underwater sounds in real-time. Dr. Peter Dugan will lead the Machine Learning Systems Integration Team at Cornell University. This group will leverage experience in applied recognition systems and focus development and integration tasks on advanced technologies for detection-classification. Dr. Dugan will leverage four senior consultants, Dr. Harold Lewis, Dr. Mark Fowler, Katie Vannicola and Dr. Rosemary Paradis; with expertise as senior professionals and faculty in the areas of Computational Intelligence, Signal Processing, Speech Processing and Neural Networks, respectively. Dr. John Zollweg, Marian Popescu, Dr. Yu Shiu, Mohammad Pourhomayoun, Katie Vanicolla and Adam Mikolajczyk will form a sub-group focused on HPC technology and software development. The Biology Team facilitates data analysis and biological direction; Dr. Christopher Clark will provide expertise and leadership in marine mammal bioacoustics, along with co-PI Dr. Sofie Van Parijs. Together, Cornell, NYU and NEFSC will perform basic and applied research, and coordinate publication efforts in engineering, biology, and systems operational research.

## Work Completed

Project funding is being provided from two sources, ONR and NFWF. A project kick off meeting was held at NYU in October 2011. Since funding from both parties was not established during FYI 2011, NYU is not obligated in this report. Cornell University utilized 30 days of funding between October and December 2011, during this time the team devoted resources to the development of the DCL hardware and software infrastructure that will be used throughout this project. Specifically the high performance-computing platform (HPC) was developed, called the HPC-Acoustic Data Accelerator, or HPC-ADA for short. The HPC-ADA was designed based on fielded systems [1-4, 6] that offer a variety of desirable attributes, specifically dynamic resource allocation and scalability. The HPC-ADA platform has been proven in three different modes of operation, ranging from stand-alone to client server based. The system is operational and supporting various contracts, Table 7, and the team has demonstrated this unit to the Board of Directors at the Cornell, Lab of Ornithology. Official kick-off started in June 2012, and the Cornell team has mobilized the initial datasets from the Stellwagen Bank National Marine Sanctuary (SBNMS)[1]. Initial development and integration of existing recognition algorithms has been performed, and preliminary results are summarized herein.

From October 2011 to December 2011, the team focused efforts on developing a DCL system that processes archival data as fast as possible. Development started by building the DeLMA-HPC software (Detection cLassificaiton for MAchine learning - High Peformance Computing). The software package was designed to utilize parallel and distributed processing for running recognition and other advanced algorithms. DeLMA software is a custom developed Matlab module which plugs into Sedna [7] and is designed using a parallel architecture[2], allowing existing algorithms to distribute to the various processing nodes with minimal changes to their structure. An advanced hardware platform (prototype) was also designed and constructed for high performance computing (HPC) applications,
Figure **1**. Various hardware components were assembled, including the rack unit, power supply modules, multi-core servers and a, custom designed, high speed network attached storage (NAS) unit. HPC-ADA uses commercially available computers that have multiple processor cores (not GPU's). The system was built as a self contained unit, using a mobile rack. HPC-ADA is capable of dynamically allocating resources to single or multiple users with a range of connectivity. Software is Matlab based, specially designed to run advanced algorithms to process audio at high speeds. Currently the HPC-ADA hardware contains 84 processors, and is scalable to larger numbers. The basic server system and I/O paths are shown in
Figure **2**. Several processing configurations were considered based on the location of the data; different combinations of distributed and local models were tried. In the archival mode, the DeLMA software allows sounds to be cached locally to the processors, reducing dependency on large network resources. To elliminate possible bottlenecks, a high speed network attached storage (NAS[3]) device

---

[1] Data recorded using various arrays of Marine Autonomous Recording Units (MARUs).
[2] Model used for archival data, a different model will likely be used for real-time audio.
[3] NAS units are commonly available, for this work our custom designed unit will leverage smart processing for later research and development efforts.

*Dugan, Clark, LeCun and Parijs*

was developed. Prototypes were created for each phase of HPC-ADA development. Full system analysis and design is beyond the scope of this research, but some simple metrics, such as throughput and resource utilizations have been gathered to baseline processing efficiency, performance shown in Table 6. It is expected that sound data will be the largest resource required to be manage by the DCL system. To address these requirements, data and control information were treated as virtual-separated network channels,
Figure **2**. The team tested out various distributed processing models by assigning detection-classification processing to each worker, which is associated with a unique ADA accelerator. The ADA computer is designed to run as if it where a single machine using one or many hardware units. In some configurations, latency between the worker and the sound source would cause the thread to time out, producing an error. These relationships were identified and elliminated through software and hardware modifications. The "head node" provided control over allocating resources to the ADA accelerators, providing scalability for the system. The "head node" manages dynamic resource allocation, allowing the operator to assign computing resources based on the detection-classificaiton job using a standard software interface. For this work, several processor configurations were tried, these are shown later in the report in Table 6 and Table 7. The software is a mixture of commercial off the shelf (COTS), and a series of custom libraries written in MATLAB 2010a-2012a. User interfaces were written using the Matlab Java tools, and a source code prototype was developed to test operation in three different modes; these include (1) standalone computer mode, (2) console HPC-ADA computer mode or (3) a client server mode, (modes 1 and 2 together), see Figure 4.

Three configurations for the DCL system were tested for functionality, as shown in Figure 4. The client application was tested to run on any machine that contained the visual runtime, and the three basic concepts of the operation were explored. Each configuration required three software packages, the standard Matlab runtime, Sedna and DeLMA. Sedna provides core algorithms for parallel processing and DeLMA contains the tools required to interface the autodetection algorithms to the distributed processing hardware for execution on multiple cpu's. Case 1: *Serial or Parallel Model*, a portable, standalone mode; capable of being used by a single computer in the field or connected to a network. Case 2[4]: *Parallel or Distributed Model*, a configuration that is able to run on the HPC server platform with access to local or distributed resources (cpu's and memory). Case 3: *Local* or *Remote Distributed Model* is the most complex setup intended to support a laboratory of users. Case 3 is designed to support centralized data and hardware resources and offers the ability to allow multiple users to connect through a local, or remote, network. This configuration would be ideal for a laboratory of users requiring accessing to a powerful collection of cpu's and large datasets. The standalone computer contains the same packages as case 1 and 2, but in addition a network connection provides communication to the server HPC-ADA machine[5]. In all cases the DeLMA package is co-resident on the client machine(s) and the HPC-ADA unit. Pull down menus in the DeLMA software (not shown) allow the user to switch between using the local computer or the network HPC-ADA machine; a performance comparison to local client machine and the HPC-ADA is shown in Table 6.

Between Januray 2012 and May 2012, the project was placed on hold due to funding delays. During this time, BRP had several projects that required detection classification work. The team adapted several existing algorithms to the HPC platform. These included: (1) the multi-stage right whale algorithm, isRAT [8], (2) a basic spectrogram correlation algorithm from xBAT using a matched image approach called the data-template [9], configured to detect fin whale, bryde's whale and

---
[4] Case 2, the HPC-ADA computer contains the same packages as case 1 with the exception that the DeLMA-*client* package is replaced by the DeLMA-*Server* package.
[5] Note: cloud based or wide sense computing applications will require an externally facing head-node. Only intranet has been tested to date.

*Dugan, Clark, LeCun and Parijs*

mechanical noise, and (3) and a new segmentation-recognition algorithm, called Acoustic Segmentaiton Recognition Algorithm (or ASR Algorithm). The new ASR algorithm was based on work from [3, 5, 6, 10, 11] and successfully used to detect various pulse train signatures including seismic air gun, minke whale, fin whale and sperm whale. A summary of the projects, which were ran using a single user, case 2 and case 3 configuration, is shown in Table 7.

From June to September 2012 the project focused on coordinating the mobilization of nearly four years of acoustic data recorded during an earlier NOPP project. Data spanned from 2006-2010 using several arrays located in SBNMS. A summary of the NOPP data is provided in Table 2. These data sets contain a variety of anthropogenic noise sources, such as commericial vessels, fishing boats, a variety of whales (fin, humpback minke, right, and sei whales) and fish (haddock and cod). The focus was to explore the data using the new ASR algorithm. Specifically the ASR algorithm was designed to interface with the HPC-ADA (Figure 1) system by grouping the output events as show in Figure 3. The network attached storage was designed to host all the continous datasets for this period, approximately 6 TB's. Some sample acoustic signatures of various pulse trains are shown in Figure 3. Three cases were tested for functionality using a subset of data from the SBNMS, 2006-2010. A dataset was constructed from portions of Table 2 that consisted of animal pulse trains and some noise events, Table 3. After running data at scale, additional noise events were added, this is referred to as the "eight-day set", Table 4. This set consisted of events from Table 3 plus additional noise samples; these proved critical for properly training and testing the ARS algorithm. The system was tested to measure throughput performance running the ARS algorithm configured for pulse train detection. For this study two versions were trained, one using exemplars from only Table 3 and the second version trained using Table 4. In total, 2429 pulse trains were taged and labeled by experienced research analysts. Exemplar sets consisted of haddock sounds, humpback social sounds, minke whale songs and an unknown signal type. For the purpose of this report, individual species were grouped together, whereby the machine learning algorithms were designed to filter out noise events. This work did not classify down to the species, which is a topic for future research. Performance for the pulse train ASR algorithm was measured using various worker configurations, see Table 6.

Various tools from Sedna were modified to allow the user to interact with the data in the diel plots, as explained in the illustration in Figure 5. For this process a montage-like browser was interfaced to the diel plot, allowing the user to select events and display these using a postage stamp view of several spectrograms. Tag labels are chosen for each event by a skilled researcher. Results of the labeling operation are stored in the log files and data tables. Tag labels used in this work are shown in Table 5. A basic configuration of the ASR algorithm was used to extract likely energy, which are displayed as the diel[6] plots to reveal daily and seasonal patterns of animal acoustic activity Figure 6. It is worth mentioning that these figures do not contain all vocally active species identfied, rather, they represent detections of pulse trains based on human observation and the ASR algorithm.

## RESULTS

In this effort it was determined[7] that the cost tradeoff for using the HPC Hardware technology is justifieable if the researchers want to run all the data and perform complex processing. Cost of the hardware will range anywhere from a full-up system (HPC rack) to multi core processor on a laptop.

---

[6] Diel plot, initial design courtesy, Melissa Soldevilla at NOAA Southeast Fisheries Science Center.
[7] All prices assume basic Mathworks license with parallel processing and other toolboxes.

5*Dugan, Clark, LeCun and Parijs*

Price range will vary, 84-core[8] stand alone appliance, will start around $75k in hardware costs to around $2k for a laptop. One major goal of this project is to investigate scalability and interoperability for advanced technologies. The ideal configuration is to allow scaling using additional server units without changing software configurations. This permits multiple computers to be added to the configuration, adjusting to the workload or data scale. A custom, high performance, network attatched server (NAS) was also developed to provide a mechanism to investigate advanced technologies for online machine learning [12, 13] and data management; such as SWARM based computing. Interoperability provides a systems construct that allows a variety of algorithms aimed at detection-classification to be used to process large datasets; for example, standard algoroithms (written in Matlab) were integrated into the HPC-ADA environment and successfully used on several projects. Interroperabilty also means that the developer does not have to make substantial modificaitons to their algorithms for integration. Speed and scale is realized using an extremely parallel architecture, allowing for efficient and fast processing, which was the primary goal for this phase. Our benchmark study shows a single desktop computer can execute a given dataset in 2 hours 55 minutes. The same dataset, takes less than 5 minutes to run on the HPC-ADA machine. Before this technology was implemented, it took roughly one month to process all the acoustic data for a single SBNMS deployment. The high performance technology can run the entire 44 months of SBNMS in less than 8 hours.

Initial development with the ASR algorithm provided several insights to working with the data from 2006-2010. First, as shown by the results in Figure 6, the scale of data used to train the algorihtms is very important. The training set will have a large impact on how well the automatic algorithm performs. While this is common knowledge, the system, and tools, may limit ones ability to access the proper scale of data. This work has allowed us to explore nearly 44 months of data, reducing many errors in an effort to provide accurate diel patterns for the animals of interest. Second, the way the algorihtms represent the data is critical to the number of objects that need to be managed by the computer. In some cases each pulse train event would generate 10-24 pulses. Likewise, seismic airgun surveys could generate millions of events over several weeks on an array of sensors, with each event capturing many features. Thus, grouping signals is helpful for reducing the number of acoustic objects that are tracked in the system. Third, detection classification over wide scales should be looked at in terms of all the acoustic objects that are captured in the desired analysis. The original work was intended to look at only minke pulse trains. After runing the algorithms on all the data, we successfully captured a variety of signals, including minke whale, humpback social sounds, haddock and cod fish sounds and a class of unidentified signals. Fourth, standard tools may not be sufficient to use on data from the HPC environment. Massive amounts of information required new data formats and new ways to access information. Intelligent organization of the data, and a well thought out plan of how users interact, is key. For example, initial manipulation of the diel plots, Figure 6, proved that older data formats were inefficient, latency in visual displays was unacceptable. New formats provided more efficient processing, data was displayed at near real time. Lastly, the technology developed herein was made possible by properly utilizing higher level programming environements like Matlab. In regards to HPC technology, it would be cost beneficial for the bioacoustic community to consider utilizing these platforms for collaborations, especially for running large scale datasets. Further attention should be placed on data standardization and tool development. High performance technologies can be ideal for distributed problems and modeling efforts. The research herein is one example, using HPC to understand diel plots for animal acoustics; other applications such as noise analysis and acoustic modeling can also be adapted to the HPC-ADA platform.

---

[8] For this research, a new Dell, hosting 64 processors was purchased for approximately $25k.

*Dugan, Clark, LeCun and Parijs*

# Figures

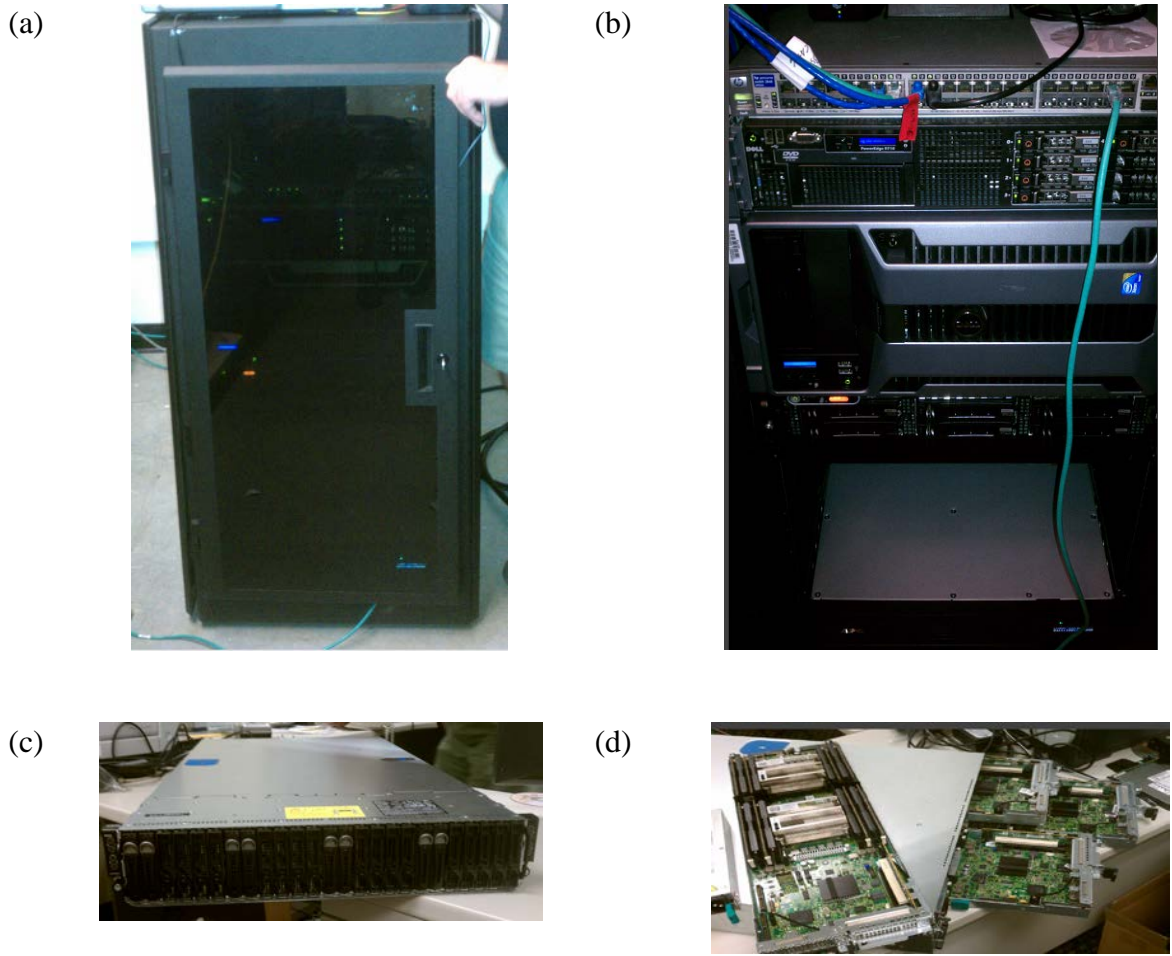

Figure 1. Photos of HPC-ADA DeLMA system. (a) System currently hosted in a portable unit, which contains all equipment to run standalone, or racked into a larger stationary server. Currently the unit has 84 processors, over 200 GB of storage and custom designed NAS unit to host many large-scale datasets. (b) View of the internal components, (HPC servers not currently in rack). (c) Dell, hosting 64 processors in a single 2U chassis. System architecture small enough to be located in a small rack, 192 GB of on board memory. (d) Rear facing HPC Dell server, 4 independent motherboards, hot swappable.

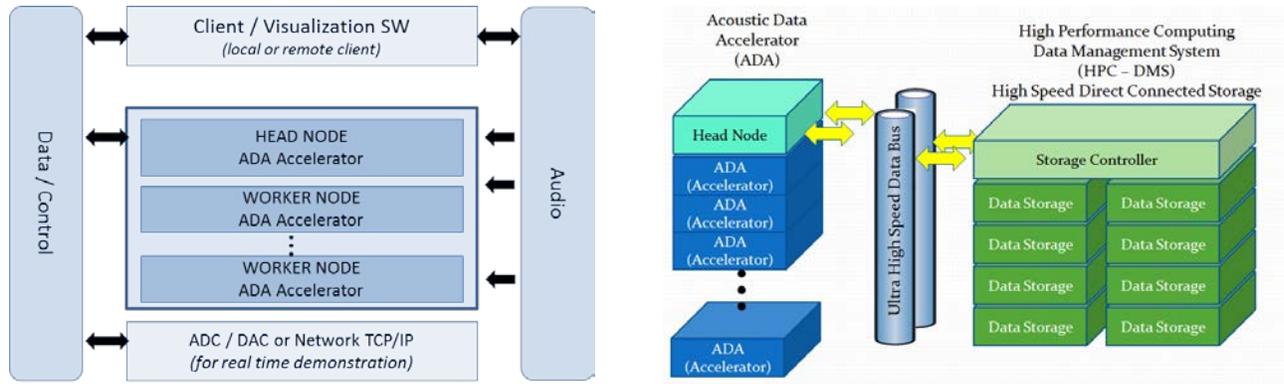

Figure 2. Communication architecture for the distributed DCL system. (*left*) The ADA computer contains a HEAD node, then a series of WORKER nodes, called accelerators. Accelerators can be a single server or a multi-board HPC server. Worker nodes added for scalability. External interfaces use a common path for data/control and one for audio. (*right*) Audio data must be accessible on a fast storage connection and visible to the client and HPC-ADA machine; data/control can be on the same storage or exchanged through a local client connection. This figure shows a custom network attached storage unit created for the research. The NAS is high performance and has the ability to utilize smart processing to explore using advanced technologies such as SWARM computing.

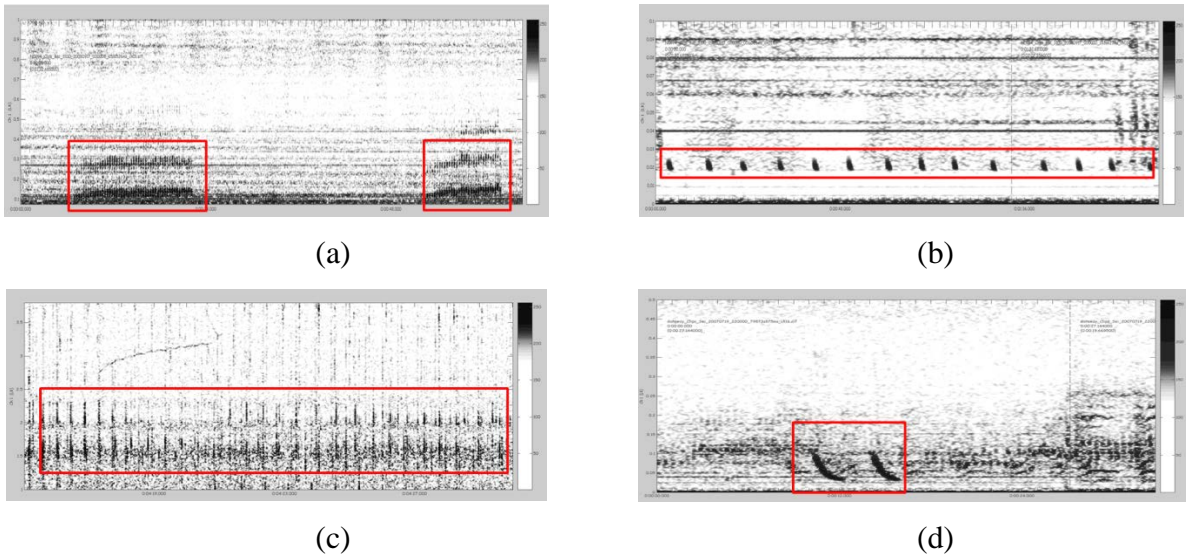

(a)                  (b)

(c)                  (d)

Figure 3. Grouped events in the spectrogram using the ASR algorithm for marine mammals. (a) Minke Whale. (b) Fin Whale. (c) Sperm Whale. (d) Sei Whale. Note: these acoustic events are not all in the SBNMS, shown as examples.

*Dugan, Clark, LeCun and Parijs*

(a)

(b)

(c)

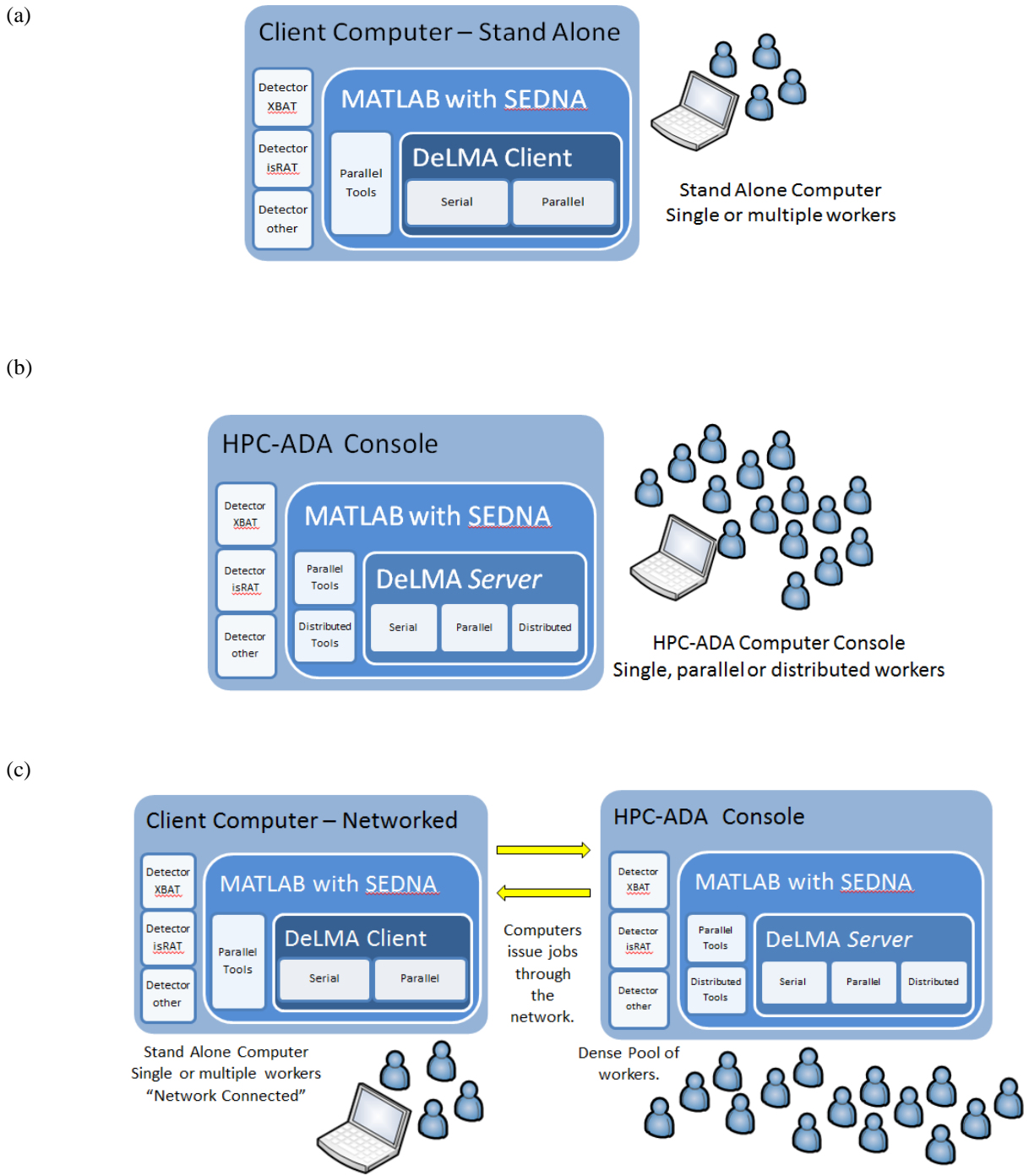

Figure 4. Figure shows three different configurations for parallel distributed computing referred to s Case 1, 2 and 3. This research uses the DeLMA software developed by Cornell University to realize each configuration. An embarrassingly parallel model allows standard detection algorithms to be added to the system, each will taking advantage of the parallel-distributed workers. (a) Case 1: Standalone mode, (b) Case 2: HPC-ADA console, serial, parallel or distributed and (c) Case 3: networked client-server model, opening up the HPC-ADA as a network appliance to other users.

*Dugan, Clark, LeCun and Parijs*

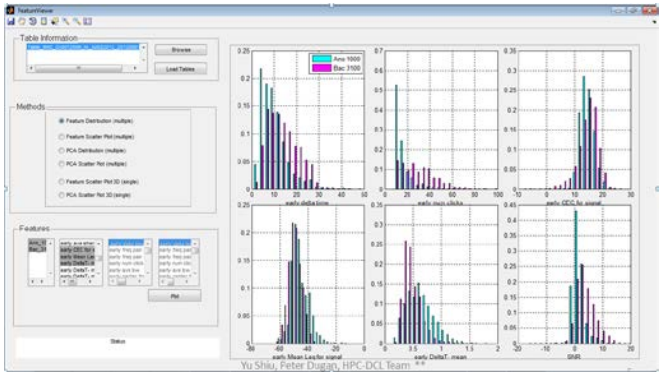
(a) Feature Distribution Plots

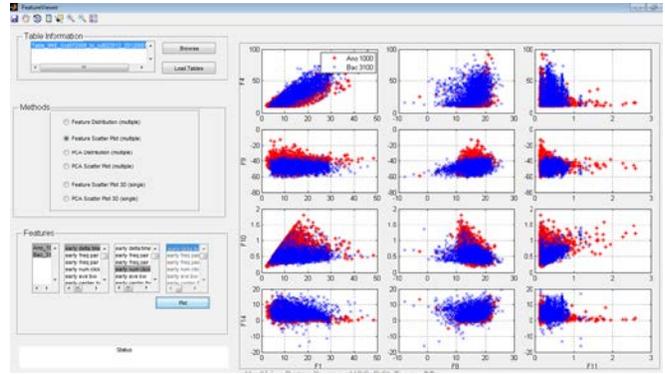
(b) Feature Scatter Plots

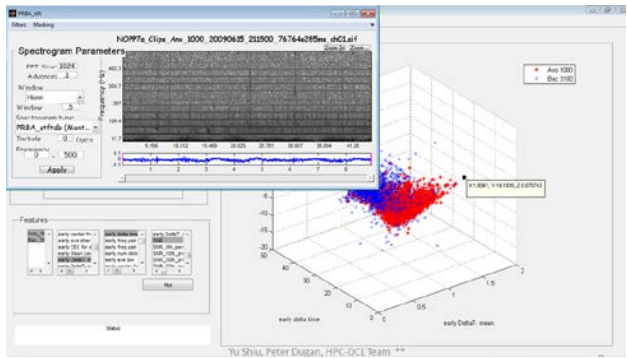
(c) Scatter plots having interoperability with signal spectrogram views.

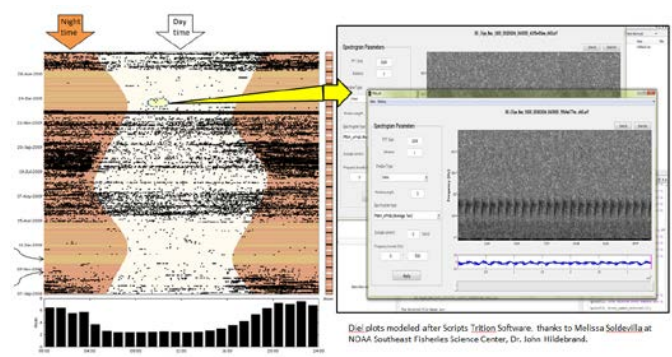
(d) Diel plots having interoperability with signal-spectrogram views.[9]

Figure 5. Sample visualization tools used to allow the human operator to interact with the data. (a) Feature distribution plot. (b) Feature scatter plot. (c) Scatter plots having interoperability with signal spectrogram views. (d) Diel plots having interoperability with diel plot views.

---

[9] Diel plot, initial design courtesy, Melissa Soldevilla at NOAA Southeast Fisheries Science Center.
10

*Dugan, Clark, LeCun and Parijs*

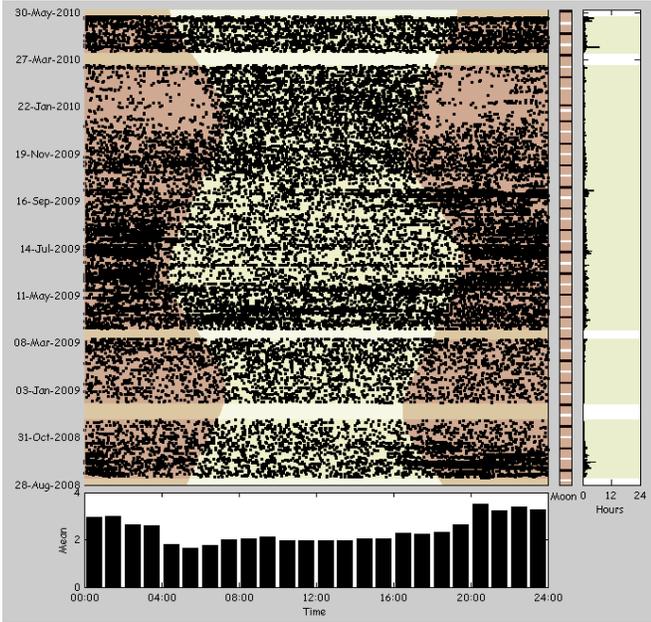

a.) Version-1, Pulse Train ASR algorithm developed using Table 3 exemplar data.

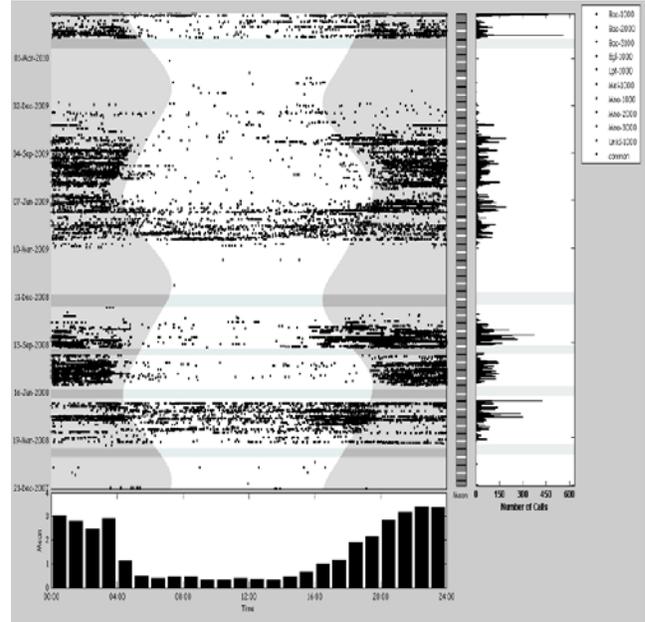

b.) Human ground truth for pulse train activity, separating animal from noise. ASR alorithm version-1 was run first.

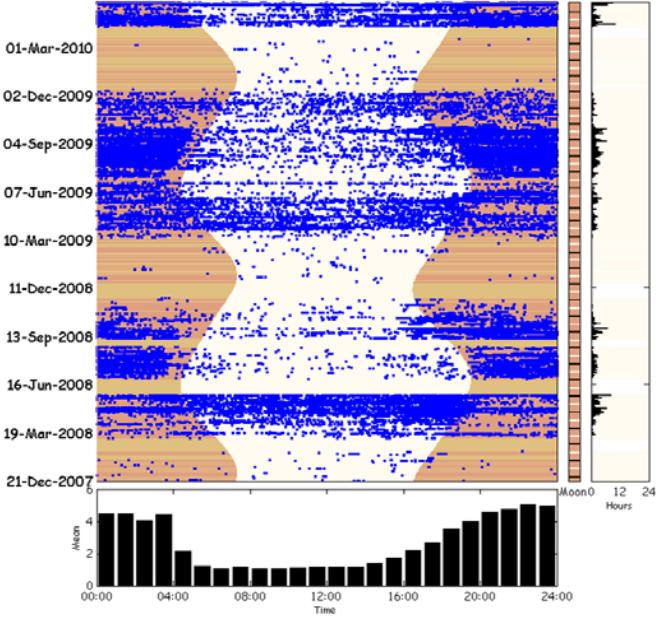

c.) ASR algorithm with training/testing data taken from Table 4, which includes exemplars from Table 3 with additional noise samples taken from all 4 years.

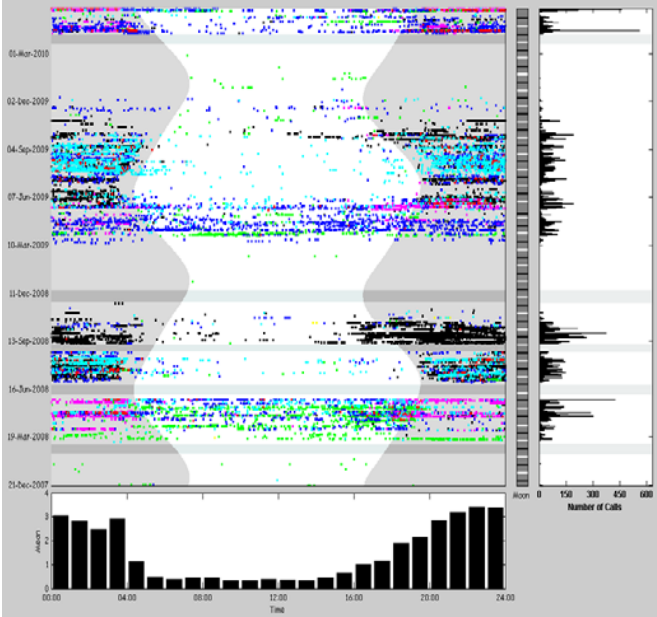

d.) Human ground truth for pulse train activity, separating different classes of pulse trains.

Figure 6. (a) Version 1 of the pulse train ASR algorithm, results looked good on a small dataset, but larger errors on continuous data. (b) Human Ground Truth using version-1 data, (c) Pulse Train detection events from ASR algorithm using version-2, error rate is much lower than versio-1, by inspection (d) color coded diel plot showing separate species pulse train activity, (note: tag labels not shown for right diel plot, these are currently being submitted for publication).

*Dugan, Clark, LeCun and Parijs*

# Tables

| NOPP-DCL Integrated Research Team (IRT) | | |
|---|---|---|
| Person | Background | Affiliation |
| **Project PI's and Component Leaders** | | |
| Dr. Peter Dugan | Engineering Scientist, PI and Project Technical Leader for Cornell BPR. Signal processing, computational intelligence and high performance computing. | Primary PI, Cornell University, Bioacoustics, Lead Scientist High Performance Computing and Recognition Team. |
| Dr. Christopher Clark | Senior Scientist, Co-PI and Lead Biologist for Cornell-BRP. Expert in marine mammal science, signal processing, acoustics and data exploration. | Co-PI, Cornell University, Director of Cornell Bioacoustics Research Program. |
| Dr. Yann LeCun | Senior Academic, Lead Scientist on convolutional neural networks and PI for NYU. Specializing in machine learning, computer vision, mobile robotics, and computational neuroscience. | PI, Silver Professor New York University, Courant Institute of Mathematical Sciences |
| Dr. Sofie Van Parijs | Lead Biological Scientist and PI for NEFSC on the NOPP datasets and ecology associated with the Stellwagen Bank National Marine Sanctuary. | PI, Lead Biologist Northeast Fisheries Science Center, Woods Hole. |
| **Biological Sciences / Ecology** | | |
| Dr. Aaron Rice | Marine biology and fish ecology. Lead Biologist for the Mass. Bay array geometries. | Science Director, Cornell University, Bioacoustics Research Program. |
| Denise Risch | Marine Biologist working primarily on minke whale, fin, humpback and North Atlantic right whale acoustic ecology and ocean noise research. | Senior Bio-acoustician, National Oceanic and Atmospheric Association, Northeast Fisheries Science Center Woods Hole, MA |
| Dr. Anne Ward | Human auditory perception for identifying cetaceans, specifically those located in the Mass. Bay area. | Senior Research Analyst, Cornell University, Bioacoustics Research Program. |
| **Technology and Engineering Sciences** | | |
| Dr. Mark Fowler | Senior Academic in Digital Signal Processing data compression and remote-distributed sensor theory/applications. | Consultant, Senior Professor Binghamton University, Electrical and Computer Engineering. |
| Dr. Harold Lewis | Senior Academic in Systems Science, specializing in soft computing, machine learning and computational intelligence. Applied math and human inference systems. | Consultant, Senior Professor Binghamton University, Systems Science. |
| Dr. Rosemary Paradis | Neural network processing, hybrid systems and applied systems. Background in applied artificial intelligent and applied computer science. | Consultant, Senior Scientist, Information Systems. |
| Dimitri Ponirakis | Noise Analysis and acoustics. Specializing in applied computing and numerical recipes for integrating sound propagation and environmental factors. | Lead Noise Analysis Engineer, Cornell University, Bioacoustics Research Program, High Performance Computing, Noise Analysis and Recognition Team. |
| Marian Popescu, MS. | Applied digital signal processing and numerical computing. Applied user interface design and advanced computer science. Design engineer for high performance computing architecture, hardware and software. | Lead Performance Engineer, Cornell University, Bioacoustics Research Program, High Performance Computing and Recognition Team. |
| Dr. John Zollweg | Applied distributed computing and software algorithm design. Background in applied math and algorithm science. | Senior Research Scientist, Cornell University |
| Dr. Yu Shiu | Digital signal processing for speech and music recognition. Algorithm design and statistical analysis and feature analysis. | Post-Doctoral Associate, Cornell University, Bioacoustics Research Program, High Performance Computing and Recognition Team. |
| Katie Vannicola, MS | Signal processing and noise removal. Applied speaker identification and speaker separation. | Consultant, Application Engineer, Air Force Research Laboratories, Rome NY. |
| Mohammad Pourhomayoun | Digital signal processing, optimization and applied math. | Cornell University, Bioacoustics Research Program, High Performance Computing and Recognition Team. |
| Adam Mikolajczyk | Applied computing systems, server hardware and computing systems. Specializing in hardware development for large scale processing and high performance computing. | Cornell University, Lab of Ornithology Computing Team. |

Table 1. Integrated Research Team (IRT) assembled for the NOPP-DCL project.

| Year | Data set | 1st full rec day | last full rec day | # Channels |
|------|----------|------------------|-------------------|------------|
| SBNMS – Pilot Study 2006 Data (MARU) | | | | |
| 2006 | Jan-March | 6-Jan | 28-Mar | 7 |
| 2006 | April-May | 30-Mar | 23-May | 9 |
| 2006 | July-Sep | 29-Jun | 21-Sep | 9 |
| 2006 | Sep-Nov | 27-Sep | 14-Oct | 9 |
| 2006 | Sep-Nov | 16-Oct | 26-Nov | 7 |
| 2006/2007 | Dec-Feb | 6-Dec | 20-Feb | 8 |
| NOPP – 2008 Data (MARU) | | | | |
| 2007/2008 | NOPP1 | 20-Dec | 16-Feb | 10 |
| 2008 | NOPP2/10ch | 8-Mar | 4-Jun | 10 |
| 2008 | NOPP3/10kHz | 26-Jun | 13-Jul | 10 |
| 2008 | NOPP3/2kHz | 14-Jul | 24-Aug | 1 |
| 2008 | NOPP4 | 9-Sep | 23-Nov | 10 |
| NOPP – 2009 Data (MARU) | | | | |
| 2008/2009 | NOPP5 | 19-Dec | 12-Mar | 10 |
| 2009 | NOPP6a | 14-Mar | 27-Mar | 1 |
| 2009 | NOPP6 | 28-Mar | 27-May | 10 |
| 2009 | NOPP7a | 29-May | 15-Jul | 2 |
| 2009 | NOPP7 | 16-Jul | 27-Jul | 10 |
| 2009 | NOPP8a | 28-Jul | 1-Oct | 1 |
| 2009 | NOPP8 | 3-Oct | 29-Nov | 10 |
| 2009 | NOPP8b | 30-Nov | 15-Dec | 1 |
| NOPP – 2010 Data (MARU) | | | | |
| 2009/2010 | NOPP9 | 16-Dec | 18-Mar | 10 |
| 2010 | NOPP10 | 8-Apr | 23-May | 10 |

Table 2. Summary of the NOPP data used on this project. Data contains nearly continuous information from 2006 to April 2010.

*Dugan, Clark, LeCun and Parijs*

| Data Set | MARU - Sensor | DAY | Number of Files | Sound Size | Source Type |
|---|---|---|---|---|---|
| Set 01 | NOPP4 | 9/17/2008 | 141 | 59.2 MB | Signal |
| | NOPP4 | 9/18/2008 | 156 | 60.0 MB | Signal |
| | NOPP4 | 9/19/2008 | 162 | 76.1 MB | Signal |
| | NOPP4 | 9/27/2008 | 69 | 17.9 MB | Signal |
| | NOPP5 | 1/15/2009 | 379 | 143.0 MB | Noise |
| Set 02 | NOPP5 | 2/15/2009 | 376 | 172.0 MB | Noise |
| | NOPP8 | 10/3/2009 | 167 | 55.3 MB | Signal |
| | NOPP8 | 10/15/2009 | 284 | 110.0 MB | Noise |
| Set 03 | NOPP6 | 5/15/2009 | 41 | 317.0 MB | Noise |
| | NOPP6a | 3/15/2009 | 267 | 93.5 MB | Noise |
| | NOPP8 | 10/27/2009 | 35 | 12.2 MB | Signal |
| Set 04 | NOPP7a | 6/15/2009 | 233 | 148.0 MB | Noise |
| | NOPP7a | 7/15/2009 | 193 | 87.8 MB | Noise |
| | NOPP8a | 8/2/2009 | 3 | 269.0 KB | Signal |
| | NOPP8a | 10/1/2009 | 76 | 28.7 MB | Signal |
| | SBNMS | 9/14/2006 | 16 | 3.0 MB | Signal |
| | SBNMS | 9/16/2006 | 1 | 463 KB | Signal |
| | SBNMS | 9/28/2006 | 2 | 877 KB | Signal |
| | SBNMS | 10/6/2006 | 4 | 1.94 MB | Signal |
| Set 05 | SBNMS | 10/7/2006 | 188 | 80.5 MB | Signal |
| | SE | 1/6/2010 | 55 | 28.1 MB | Signal |

Table 3. Subset of NOPP data used to test-train Version-1, pulse-train Acoustic Segmentation Recognition (ASR) Algorithm.

| Data Set | Sound Duration | Sample Rate (Fs) | Number of Present Hand Truth Animal Calls |
|---|---|---|---|
| Set 01 | 26 Hrs - 01 Mins - 40 sec | 2000 Hz | 1141m |
| Set 02 | 24 Hrs - 39 Mins - 01 sec | 2000 Hz | 432 |
| Set 03 | 30 Hrs - 51 Mins - 26 sec | 2000 Hz | 57 |
| Set 04 | 19 Hrs - 47 Mins - 11 sec | 2000 Hz | 194 |
| Set 05 | 07 Hrs - 54 Mins - 55 sec | 2000 Hz | 605 |
| Set 06 | 00 Hrs - 29 Mins - 42 sec | 2000 Hz | 0 |
| Set 07 | 01 Hrs - 50 Mins - 07 sec | 2000 Hz | 0 |
| Set 08 | 02 Hrs - 27 Mins - 50 sec | 2000 Hz | 0 |
| Total | 114 Hrs -10 Mins - 48 sec | 2000 Hz | 2429 |

Table 4. Hand truth call table, summary of marine mammal activity for pulse train signals summarized from the test and training. Sets 6, 7 and 8 contain only noise events taken across the entire SBNMS four year deployment. Data above is comprised of the additional noise and exemplars in Table 3. Data shown above provided a more robust set for buildingVersion-2 of the ASR algorithm.

*Dugan, Clark, LeCun and Parijs*

| Tag | Tag Description |
|---|---|
| Bac_1000 | Possible minke whale, pulse train which could be made by minke but not definite |
| Bac_3100 | Definite minke whale |
| Ano_3100 | Noise |
| Hdd_3100 | Hard drive, device noise |
| Mel_1000 | Haddock |
| Mno_1000 | Moan pulse train (likely source: humpback) |
| Mno_2000 | Humpback song |
| Egl_1000 | Right whale |
| Lpt_1000 | Low frequency pulse train (source: unknown) |
| Mno_3000 | Pulse train made by humpback |
| Unid_1000 | Unidentified source |

Table 5. Tag label and description.

**Dell Desktop versus HPC-ADA**
**Benchmark Detection Runtime Performance**

| Dell Desktop Work Station | | | |
|---|---|---|---|
| Number Cores | Processor | Elapsed time Hrs:MM:Sec | Number of Detected and Classified Events |
| 1 | Intel Xeon X5482 @ 3.2 GHz | 2:55:12 | 16863 |
| 4 | Intel Xeon X5482 @ 3.2 GHz | 1:28:28 | 16863 |
| HPC-ADA | | | |
| 4 | Intel Xeon X5650 @ 2.67 GHz | 0:44:48 | 16863 |
| 12 | Intel Xeon X5650 @ 2.67 GHz | 0:23:24 | 16863 |
| 22 (10 Virtual) | Intel Xeon X5650 @ 2.67 GHz | 0:19:12 | 16863 |
| 64 | Intel Xeon X5650 @ 2.67 GHz | < 0:5:0 | 16863 [10] |

Data Size = 2 GB; Sound Size = 114Hours, 10 Minutes, 48 Seconds (Continuous Recording); Sample Rate (Fs) = 2000 Hz

Table 6. Example, Detector-Classifier runtime performance. Comparing various worker configurations when running the pulse train detector used the animal vs. noise example. Results gathered by running on (1) standard desktop computer, Dell 8500, (2) HPC-ADA server.

---

[10] 64 core, projected results using the DeLMA cloud computing HPC architecture, using Dell Power Edge 610T Server, the new DELL c6220 4 Node Server, estimated speeds are expected to be less than 5 minutes, since the processors host bus architecture is faster than the PE-610T server.

*Dugan, Clark, LeCun and Parijs*

## IMPACT/APPLICATIONS

Currently the authors do not know of anyone that has successfully integrated HPC technology for doing advanced detection-classificaiton for marine mammals and passive acoustic archival data using a multiple configuration (see Case 1, 2, and 3), Figure 4. These onfigurations may serve as a starting point for various reseach and development envornments, such as at Naval Processing Labs or other university centers that wish to host HPC technologies for passive acoustic research.

## RELATED PROJECTS

The HPC-ADA computer was developed using pre-spending funding available from ONR. The initial build, test and development of the DeLMA software occurred duing October to November, 2011. The HPC-ADA hardware and distributed software was erected during the project start, June, 2012. Between October 2011 to June 2012, several projects were executed using this technology. We used a spiral development process, which consisted of a test, build, integration and execution cycle. This process allowed us to execute on various projects while the system was being developed (Table 7). Integration of this work consisted of adapting various detector classifiers that were currently being run using other MATLAB tools (xbat and isRAT) to the DeLMA software. During this cycle, we had two multi-core servers available to perform testing and runtime.

| Project | Amount of Data Processed, Including Multi-Channel Data. | # Channels | Sample Rate | # Cores Used for Processing | Type of Processing |
|---|---|---|---|---|---|
| Excellerate | ~ 150 days / 0.63 TB | 19 | 2 kHz | 12 | Massechessets Bay, standard north atlantic right whale NARW detectors and fin whale detectors. Detector migrated from existing tools. NARW |
| GoMex | ~ 4.3 years / 2.6 TB | 1 | 20 kHz | 24 | Gulf of Mexico, developed sperm whale detector. Work is not published and in preperation. |
| CAIRN | ~ 60 days / 0.7 TB | 5 | 16 kHz | 24 | Baffin Bay, Seismic detection and feature extraction. System managed over 8 milllion feature points for auto detection and noise analysis. |
| NOPP – DCL (this project see Table 2) | ~3.5 years/ 6 TB | 1-10 | 2kHz, 10kHz | 38-64 | Stellwagen Bank National Marine Sanctuary, development of pulse train detection, animal vs noise (see example in this report). Minke, Humpback (social sounds), Haddock, Fin detection, Right Whale Detection. |
| Mass CEC | ~ 180 days / 0.4 TB | 6 | 2kHz | 12 | Massechessets Bay, run NARW, Fin and Minke whale detectors. |
| Gulf of Maine | ~3 years / 0.6 TB | 1 | 2kHz | 12 | Maine area, run NARW, Fin and Minke whale detectors. |

Table 7. Projects that have used a combination of the DeLMA software and the HPC-ADA hardware during the spiral development phases.

16*Dugan, Clark, LeCun and Parijs*

**PUBLICATIONS**

P.J. Dugan, D.W. Ponirakis, J.A. Zollweg, M.S. Pitzrick, J.L. Morano, A.M. Warde, A.N. Rice and C.W. Clark, "SEDNA - Bioacoustic Analysis Toolbox Matlab Platform to Support High Performance Computing, Noise Analysis, Event Detection and Event Modeling." *IEEE Explore*, vol. OCEANS-11, Kona, Hawaii.

**PATENTS**

Dugan, P, J., Clark, C.W., Ponirakis, D.W., Pitzrick, Rice, A., M.S., Zollweg, J.A., "System and Methods of Acoustic Monitoring", Provisional Patent Application, Case No. 5643-01-US, September 29, 2012.